\documentclass[aps,prl,reprint,superscriptaddress,twocolumn,nofootinbib]{revtex4-1}
\pdfoutput=1

\usepackage{amsmath,amssymb,amsthm,mathrsfs,bbm}
\usepackage{graphicx}
\usepackage[pdftex,colorlinks=true]{hyperref}
\usepackage{enumitem}

\newcommand{\be}{\begin{equation}}
\newcommand{\ee}{\end{equation}}
\newcommand{\bea}{\begin{eqnarray}}
\newcommand{\eea}{\end{eqnarray}}

\begin{document}

\title{Driven black holes: from Kolmogorov scaling to turbulent wakes}
\author{Tomas Andrade}
\affiliation{Departament de F\'isica Qu\`antica i Astrof\'isica, Institut de Ci\`encies del Cosmos, Universitat de Barcelona, Mart\' i Franqu\`es 1, E-08028 Barcelona, Spain.}

\author{Christiana Pantelidou}
\affiliation{Centre for Particle Theory and Department of Mathematical Sciences, Durham University, Durham, DH1 3LE, UK}
\affiliation{School of Mathematics, Trinity College Dublin, Dublin 2, Ireland}

\author{Julian Sonner}
\affiliation{Department of Theoretical Physics, University of Geneva, 24 quai Ernest-Ansermet, 1211 Gen\`eve 4, Suisse}

\author{Benjamin Withers}
\affiliation{Department of Theoretical Physics, University of Geneva, 24 quai Ernest-Ansermet, 1211 Gen\`eve 4, Suisse}
\affiliation{Mathematical Sciences and STAG Research Centre,University of Southampton, Highfield, Southampton SO17 1BJ, UK}

\date{\today}

\begin{abstract}
General relativity governs the nonlinear dynamics of spacetime, including black holes and their event horizons.
We demonstrate that forced black hole horizons exhibit statistically steady turbulent spacetime dynamics consistent with Kolmogorov's theory of 1941. As a proof of principle we focus on black holes in asymptotically anti-de Sitter spacetimes in a large number of dimensions, where greater analytic control is gained.
We also demonstrate that tidal deformations of the horizon induce turbulent dynamics. When set in motion relative to the horizon a deformation develops a turbulent spacetime wake, indicating that turbulent spacetime dynamics may play a role in binary mergers and other strong-field phenomena.
\end{abstract}

\maketitle

\section{Introduction}
The recent observations of black hole mergers \cite{GW150914} add to the increasing evidence that black holes exist in nature. Whilst at early times black hole mergers are described by perturbative post-Newtonian physics \cite{Blanchet2002}, and at late times are described by Kerr plus a handful of the longest lived quasinormal modes \cite{PhysRevLett.72.3297}, at intermediate times there is the exciting possibility that we are witnessing the full nonlinearity of general relativity \cite{Pretorius:2005gq}. 

A remarkably universal consequence of non-linear dynamics is turbulence, seen across a wide variety of systems that exhibit fluid-like behavior at the largest scales. The dynamics of the chaotic cascade of vortices across scales washes out all memory of how those vortices were created in the first place, resulting in universal characteristics. Turbulence is arguably universality \emph{par excellence}, the consequences of which are witnessed in all corners of nature from galaxy formation to atmospheric dynamics to a cup of tea.

In this paper we examine the possibility that the dynamics of black holes exhibit similar universal features to fluids undergoing turbulent cascades.  As is well known, the physics of black hole horizons and the dynamics of fluids are closely related, depending on the context. It was appreciated early on that black-hole horizons can be thought of as fluid membranes \cite{damour,Price:1986yy,membranebook}. More recently, a one-to-one map between near-equilibrium black holes solutions in asymptotically anti-de Sitter (AdS) spacetimes and the solutions to conformal hydrodynamics with particular transport coefficients  has been established \cite{Policastro:2001yc, Policastro:2002se, Bhattacharyya:2008jc}.
Recent connections between near horizon dynamics and incompressible Navier Stokes equations were studied in \cite{Bredberg:2011jq}, and moreover it has recently been established that the Stokes equations govern transport properties of inhomogeneous horizons \cite{Donos:2015gia}.
We may therefore hope that the remarkable universality observed in turbulent cascades is also seen in the dynamics of black holes. This is our present goal.

The most widely celebrated results on the universality of turbulent cascades are captured by Kolmogorov's theory of 1941 \cite{K41a, K41b} (K41).\footnote{English translations: \cite{K41a_en, K41b_en}.} Under similarity hypotheses for homogeneous isotropic turbulence, the statistical distributions of the velocity field in the inertial range depend only on the rate of transfer of kinetic energy within the cascade, $\varepsilon$. Dimensional analysis then reveals that the two-point functions of the velocity field $\vec{v}$ in momentum space -- here written in terms of the kinetic energy spectrum, $E(k)$ -- takes a simple scaling form,
\be
E(k) \equiv \partial_k \int_{|k'|\leq k} \frac{d^dk'}{(2\pi)^d} |\vec{v}_{k'}|^2 = C \varepsilon^{2/3} k^{-5/3}, \label{K41spectrum}
\ee
while for higher $n$-point functions of the velocity field in position space, arranged into longitudinal structure functions,
\be
S_n(r)\equiv\langle \left|(\vec{v}(\vec{x}+r \hat{y})-\vec{v}(\vec{x}))\cdot \hat{y}\right|^n \rangle = C_n \varepsilon^{n/3}r^{n/3}.\label{K41structure}
\ee
Here the angle brackets denote statistical averages.
These results apply in any number of dimensions, so long as the underlying details of the dynamics meet the nontrivial test of the similarity hypotheses.
We shall demonstrate that certain driven black hole spacetimes exhibit these universal features, namely we shall numerically demonstrate that \eqref{K41spectrum} holds over 1-2 decades in momentum space, and \eqref{K41structure} holds up to $n=10$ over a decade.

These properties are demonstrated for horizons with dynamics restricted to two spatial dimensions (i.e. spacetimes with nontrivial dynamics in 3+1 dimensions). In two spatial dimensions the underlying dynamics which realise the scaling hypotheses and lead to \eqref{K41spectrum} take the form of an \emph{inverse} cascade, with power moving from a driving scale at some high-$k$ to low-$k$ over time. In two dimensions there is also a \emph{direct} cascade which moves from the driving-$k$ to the UV \cite{Kraichnan}, and whilst we do see this occur, its properties will not be the focus of our work.

We focus on the simplest possible setting in which we can explore such turbulent horizons, as a proof of principle of the above properties. To this end we restrict our attention to horizons with planar (or toroidal) topology, rather than spherical. In asymptotically flat spacetimes planar horizons present an unstable starting point because of the Gregory-Laflamme instability \cite{Gregory:1993vy}, and are therefore unnatural objects to consider in our present goal of studying the dynamical response to a forcing term.  We focus instead on asymptotically AdS spacetimes where planar horizons are intrinsically stable. Such black holes are also of interest as models of strongly interacting many body systems through the AdS/CFT correspondence \cite{Maldacena:1997re}. Due to the universality of the mechanism of turbulence we may hope that our proof of principle examples shed light onto the universal dynamics of black holes in general, including those in asymptotically flat spacetimes.

In obtaining our results we do not utilise a hydrodynamic expansion, i.e. we do not carry out perturbation theory in gradients. Without this technical crutch, the connection between black holes and fluid-like dynamics is not readily apparent from  the Einstein equations. However, the connection is once more seen if one treats the inverse spacetime dimension $1/D$ as a perturbative parameter according to the seminal constructions of \cite{Emparan:2015hwa, Bhattacharyya:2015dva, Emparan:2015gva, Emparan:2016sjk}. 

At large $D$ there is a separation of scales between the black hole size, $r_0$ and the region occupied by a nontrivial gravitational potential, $r_0/D$. A separation of scales signals an effective theory, which can be constructed by analytically solving the integrals for radial evolution. What remains is a set of constraint equations in $D-1$ dimensions. It is these constraints that resemble fluid-dynamics equations. Crucially, however, even though they are perturbative in $1/D$, they are exact in gradients.
Therefore, far from a mere technical simplification, the $1/D$ expansion allows us to directly connect black holes with the turbulent behaviour of a class of fluid-like equations that are exact in gradients. We thus note that these equations may be valuable in their own right as a natural candidate for the study of turbulent behaviour, as compared to those obtained in an arbitrarily truncated gradient expansion (such as the Navier-Stokes equations). We return to this point in the discussion. 

The driving we consider is obtained by adding forcing terms directly to the black hole equations, $F_i(t,\vec{x})$, in a way that injects vorticity consistent with the requirements of statistical homogeneity and isotropy. Due to the nature of our particular setup, we are also afforded the opportunity to drive the horizon fluid by turning on a deformation to the gravitational potential at the boundary of AdS, $\gamma_{tt}(t,\vec{x})$, using a generalisation of the large-$D$ equations derived in \cite{Andrade:2018zeb}. 

Previous work has explored decaying turbulent dynamics of black holes that result after starting from unstable initial conditions \cite{Adams:2012pj,Adams:2013vsa, Chesler:2013lia, Rozali:2017bll}, where \cite{Rozali:2017bll} also utilised the large $D$ expansion as we do here. To distinguish the work we present here, we do not require starting with unstable initial data, and the driving allows us to achieve a quasi-stationary\footnote{It is quasi-stationary because we work at finite volume, without the customary addition of a friction term that dissipates at large scales.}  turbulent regime that can be compared with the predictions of K41. Motivated by the holographic connection there has also been a focus on turbulence in $2+1$ conformal hydrodynamics \cite{Carrasco:2012nf, Green:2013zba, Westernacher-Schneider:2015gfa, Westernacher-Schneider:2017snn} where analyses of the energy spectrum and comparisons to \eqref{K41spectrum} are made. Quasi-normal mode resonances of a rapidly spinning Kerr black hole have been argued to result in a phenomenon resembling an inverse turbulent cascade in 2+1 dimensions \cite{Yang:2014tla}.

\section{Large D black hole dynamics in AdS}
We shall now record the most salient points regarding the effective theory that describes the dynamics of black branes in asymptotically AdS$_D$ spacetimes at large $D$, derived in \cite{Emparan:2015hwa, Bhattacharyya:2015dva, Emparan:2015gva, Emparan:2016sjk} (see also \cite{Dandekar:2016jrp}).
In taking the large $D$ limit we focus on the near horizon region of the black brane, so that the resulting theory 
effectively describes how the near horizon deformations evolve in time. Furthermore, a mathematical simplification 
arises that allows us to solve the constraints in the radial direction, so that the basic variables can be 
readily related to the energy and momentum density of a fluid. 
This effective theory was later extended in \cite{Andrade:2018zeb} by introducing 
a general class of boundary conditions which induce changes in the gravitational potential on the 
boundary of the near-horizon region. These map to sources for the dual stress-energy tensor in the 
dual picture. Among the class of deformations derived in \cite{Andrade:2018zeb}, here we will only consider the one corresponding to adding a source for the energy density.

Our action is simply Einstein-Hilbert with negative cosmological constant in $D$ dimensions
\begin{equation}
	S = \frac{1}{16\pi G_N}\int d^D X \sqrt{-G} (R(G) - 2 \Lambda )
\end{equation}
where $\Lambda = -(D-1)(D-2)/2$.
We choose a coordinate system adapted to the black brane; we split the space-time coordinate $X^\alpha$ into
$X^\alpha = \{t, r, x^i,  y^a\}$, where $r$ is the coordinate transverse to the brane, which also plays the role of the holographic coordinate,
$t$, $x^i$ and $y^a$ are the coordinates along the black brane ($i = 1..p$, $a=1..\tilde{n}$), so that  $D = \tilde{n}+p+2$. The reason for the split between $x^i$ and $y^a$ is that we restrict only to dynamics in a subset $p$ of the boundary spatial directions, in other words, we dimensionally reduce on an $\tilde{n}$-torus and keep only the zero modes.

As in \cite{Emparan:2015hwa}, we take the large $D$ limit by taking $\tilde{n} \to \infty$ keeping $p$ fixed. 
This is facilitated by choosing a metric ansatz of the form
\begin{align}
\label{G ansatz}
 G_{\alpha \beta} dX^\alpha d X^\beta &= 2 dt dr + r^2 G_\perp d \vec y ^2 \\ 
\nonumber
	& + r^2( - A dt^2 - 2 A_i dt dx + G_{ij} dx^i dx^j  ) , 
\end{align}
\noindent where $A$, $A_i$, $G_{ij}$, $G_\perp$ are functions of $\{t, r, x^i \}$.
As shown in \cite{Emparan:2015gva, Emparan:2016sjk, Andrade:2018zeb}, it is consistent to solve the Einstein equations with the following ansatz as a perturbative expansion in $1/\tilde{n}$
\begin{align}
	A &= \left( 1 - \frac{a(t,x^i)}{R} \right) - \gamma_{tt}(t,x^i)\frac{1}{\tilde{n}}  + O(\tilde{n}^{-2}), \\
	A_i &= \frac{p_i (t, x^j)}{R} \frac{1}{\tilde{n}} + O(\tilde{n}^{-2}),  \\
	G_{ij} & = \delta_{ij} \frac{1}{\tilde{n}} + O(\tilde{n}^{-2}),  \\
	G_\perp & = \frac{1}{\tilde{n}} + O(\tilde{n}^{-3}),
\end{align}
\noindent where $R = r^{\tilde{n}}$. Here, $a$, $p_i$ are the energy and momentum density of the black brane, while $\gamma_{tt}$ is a deformation of the AdS-boundary metric. As such, we can think of it as providing an external gravitational potential, or providing an adjustment to the gravitational environment in which the black hole lives.\footnote{In the language of 
AdS/CFT, this deformation corresponds to introducing an explicit source for the energy density of the dual theory.} 
The ansatz \eqref{G ansatz} can be thought of as a radial foliation of spacetime. As a consequence, the Einstein equations 
split into evolution equations in the radial direction and constraints on $r = {\rm const}$ surfaces. The radial equations can be solved 
order by order in $1/\tilde{n}$, so we are left with the constraints. At leading order in $1/\tilde{n}$, these are
\begin{align}
	(\partial_t - \partial_i \partial^i ) a + \partial_i p^i &= 0 \label{evo_a}\\
	(\partial_t - \partial_j \partial^j ) p_i + \partial_i a + \partial_j \left( \frac{p_i p^j}{a} \right)&= F_i \label{evo_p}
\end{align}
where in the presence of the boundary metric deformation we have $F_i = \frac{a}{2} \partial_i \gamma_{tt}$ and where indices are raised and lowered with the flat metric $\delta_{ij}$.
Equations \eqref{evo_a},\eqref{evo_p} correspond to conservation equations associated to time and spatial translation invariance, and behave in accordance with expectations for a viscous fluid, complete with sound and shear-diffusion modes as detailed in \cite{Emparan:2016sjk}. These equations will serve as the basis of our analysis. 

All that remains is to provide the relations between these gravitational variables ($a,p_i$), in which we carry out numerical evolution, and a set of fluid variables: the fluid energy density is $a$, whilst the velocity is $v^i = (p^i - \partial_i a)/a$, see \cite{Andrade:2018zeb} for details. Note that this large $D$ limit results in a set of non-relativistic equations.
\section{Kolmogorov scaling}
In this section we present the key results of our work, that when appropriately driven by a homogeneous and isotropic forcing function, the large-D equations of general relativity in AdS \eqref{evo_a}-\eqref{evo_p} exhibit turbulent behaviour consistent with the predictions of K41 theory. We verify the agreement with K41 by carrying out 256 independent random realisations and using them to compute statistical properties of the velocity field, $v^i$. 

To achieve the conditions required for K41, namely homogeneity, isotropy and driving, we replace $F_i$ on the right hand side of \eqref{evo_p} by an explicit forcing function, instead of utilising $\gamma_{tt}$, which we set to zero. While $\gamma_{tt}$ \emph{is} such a forcing function, it appears as the derivative of a scalar and cannot be directly used to supply a source of vorticity.\footnote{We note that it can be used to \emph{stir} the fluid, however this turns out to be far less efficient in practise, and is dominated by frequencies associated to the stirring, at least for the examples we have considered.} We shall consider flows in the presence of nontrivial $\gamma_{tt}$ later. We adopt periodic boundary conditions for a torus of size $L\times L$. Details of the choice of $F_i$, the numerical methods used and their implementation are given in the appendix.

\begin{figure*}[t]
\begin{center}
\includegraphics[width=0.95\columnwidth]{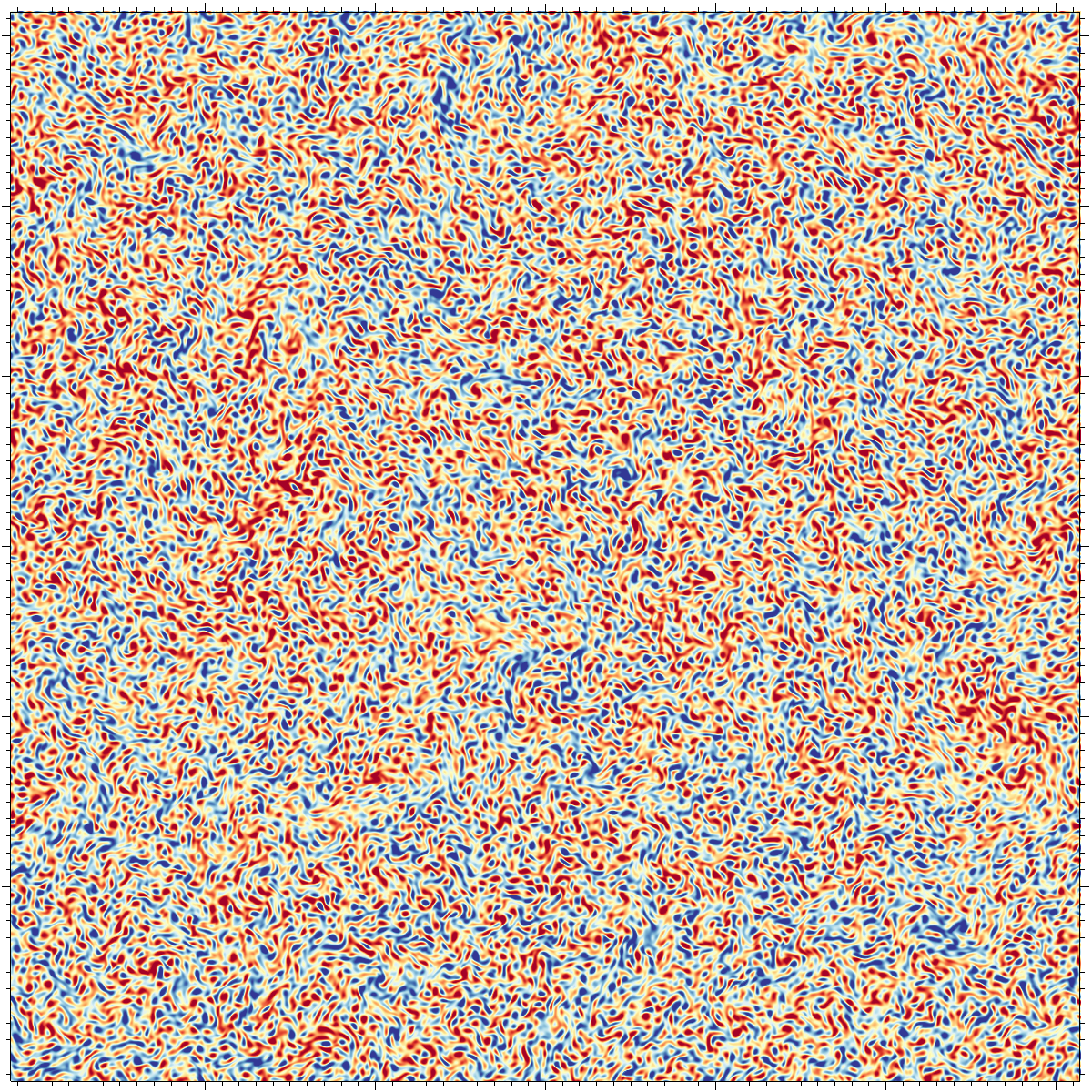}\hskip0.5cm\includegraphics[width=0.95\columnwidth]{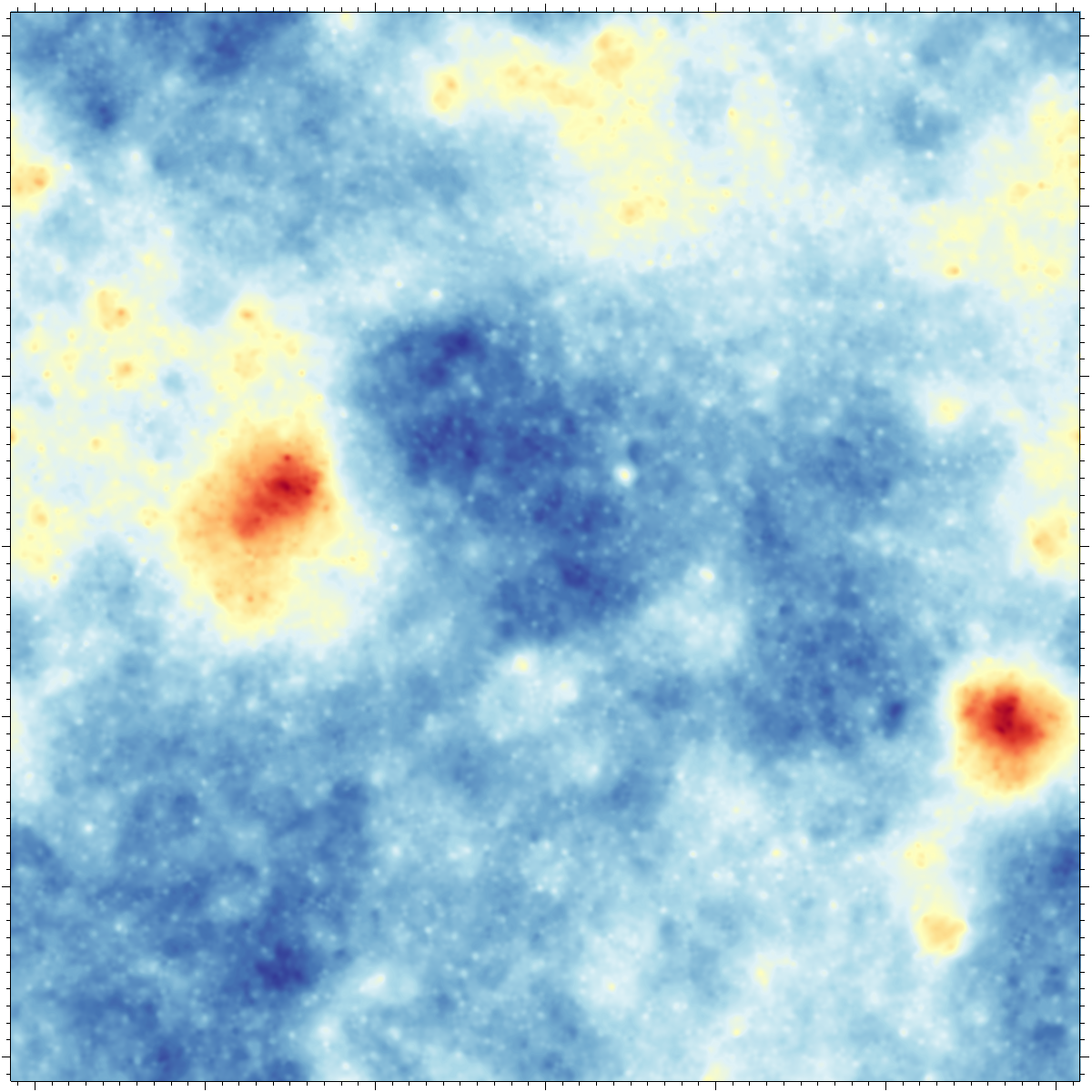}
\caption{{\bf Left panel}: Snapshot of vorticity for homogenous isotropic $F_i$-driven turbulence of a black brane horizon at $t=24L$. 
The time evolution resembles that of a vortex liquid flowing coherently as part of larger vortical structures, not apparent in this snapshot. The correlation functions of this velocity field are consistent with K41 scaling relations.\label{fig:snapshot} {\bf Right panel}: Snapshot of energy density, $a$, for homogenous isotropic $F_i$-driven turbulence of a black brane horizon at $t=24L$, for the evolution in figure \ref{fig:snapshot} (left). Inhomogeneities have amplitudes of order $10\%$ of the local energy density. At this time in the evolution, energy at high scales is beginning to `condense' into a vortex lattice at large scales, dictated by the finite size $L$ of the setup.\label{fig:asnap}}

\includegraphics[width=0.98\columnwidth]{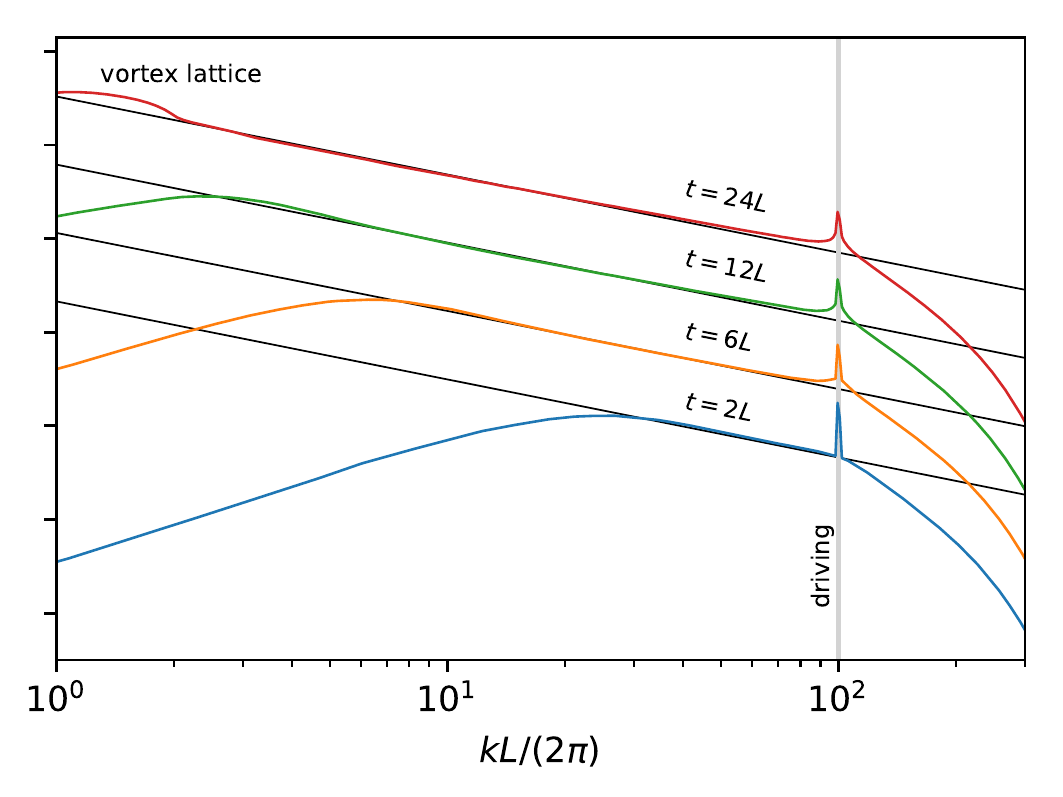} \includegraphics[width=1.05\columnwidth]{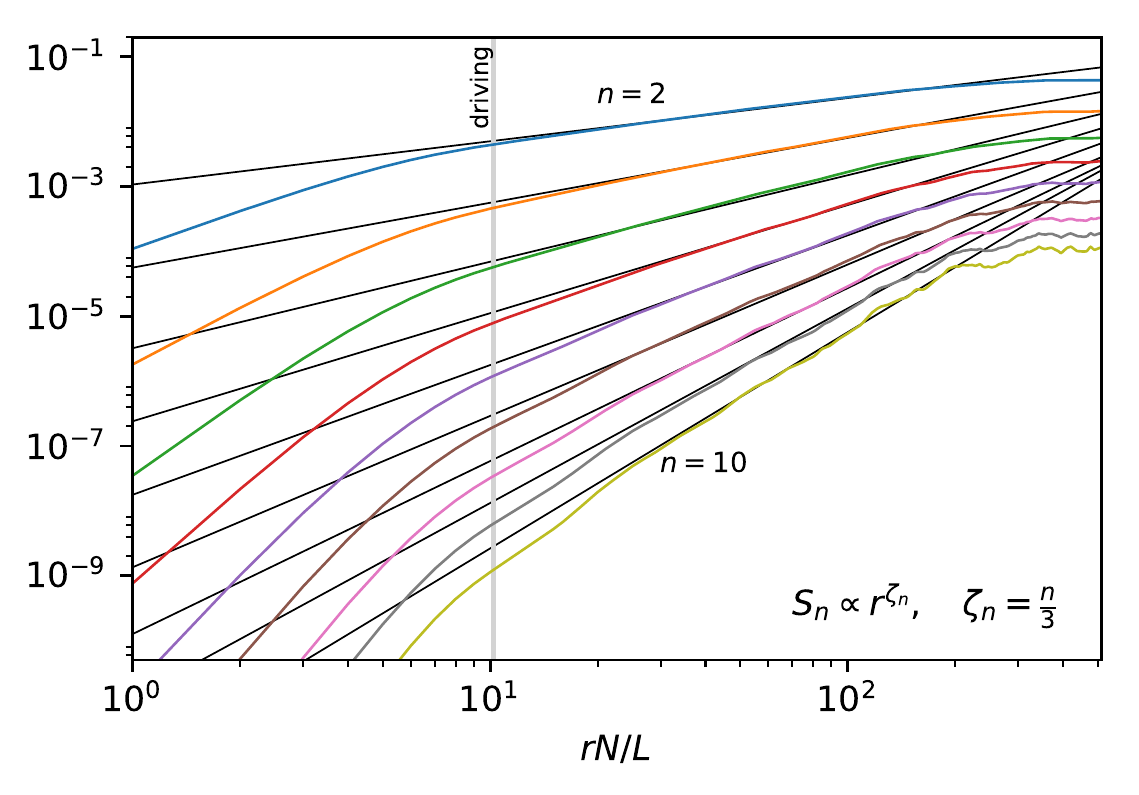}
\caption{{\bf Left panel}: Log of the kinetic energy power spectra of a turbulent black brane horizon, averaged over 256 realisations of the $F_i$-driving. The normalisation is arbitrary, with the spectra shown at four different times given different normalisations for clarity of presentation. The inverse cascade is evident as power moves to larger scales over time. The vertical grey line marks the driving scale. The black diagonal lines show the K41 scaling \eqref{K41spectrum}, $k^{-5/3}$, as a guide to the eye. At late times power accumulates at the largest scales on the torus indicating the formation of a vortex lattice structure.\label{fig:power} {\bf Right panel}: Longitudinal structure functions of a turbulent black brane horizon, averaged over 256 realisations of the $F_i$-driving, shown at $t=14L$. The straight black lines show the K41 scaling relations \eqref{K41structure} for the longitudinal structure functions up to $n=10$. The vertical grey line indicates the wavelength of the driving.
\label{fig:structure}}
\end{center}
\end{figure*}

To acquaint the reader, we first illustrate the pattern of vorticity, $\omega = \epsilon^{ij} \partial_i v_j$, obtained during a single realisation at late times in figure \ref{fig:snapshot} (left). The time-dependence of this picture resembles that of a liquid of vortices moving in larger coherent structures. Larger structures can be seen in this picture in accordance with power beginning to accumulate on the largest scales available, i.e. the torus size. This part of the spectrum, subject to finite-size effects, is expected to lie outside the inertial range. For comparison the energy density $a$ is shown in figure \ref{fig:asnap} (right) where the large-scale structures of spacetime are more easily seen.

Next we present the kinetic energy spectrum in figure \ref{fig:power} (left), averaged over all realisations. The figure is presented with arbitrary normalisation on a log-log axis, with the driving scale $|k|L/(2\pi)=100$ indicated by the vertical line and an excess of power there. The power-spectra are consistent with the K41 prediction \eqref{K41spectrum} as shown by the indicated $k^{-5/3}$ lines. 

The inertial range is seen to grow over time, from the driving scale downwards, i.e. an inverse cascade. The inertial range approaches two decades-worth of K41 scaling before meeting the finite size of the box. At this time power begins to accumulate at low $k$ and the scaling is destroyed there. Our simulations thus demonstrate quasi-stationary turbulence. Interestingly, since we are working on a torus, the structure begins to resemble that of a square vortex-antivortex lattice at late times, but continually fed from high $k$. This phenomenon also goes by the term `energy condensation' \cite{PhysRevLett.99.084501}.

Next we demonstrate that the longitudinal structure functions \eqref{K41structure} are also seen in our simulations, from $n=2$ up to $n=10$ in figure \ref{fig:structure} (right). For this position-space calculation we pick a single direction in which we compute velocity differences, $\hat{x}$, average over each row labelled by $y$, and then over 256 realisations. The consistency with K41 is indicative of the absence of intermittency corrections; we are driving the system homogeneously and not providing any window of opportunity for intermittent laminar flow to develop.

\begin{figure}[h!]
\begin{center}
\includegraphics[width=\linewidth]{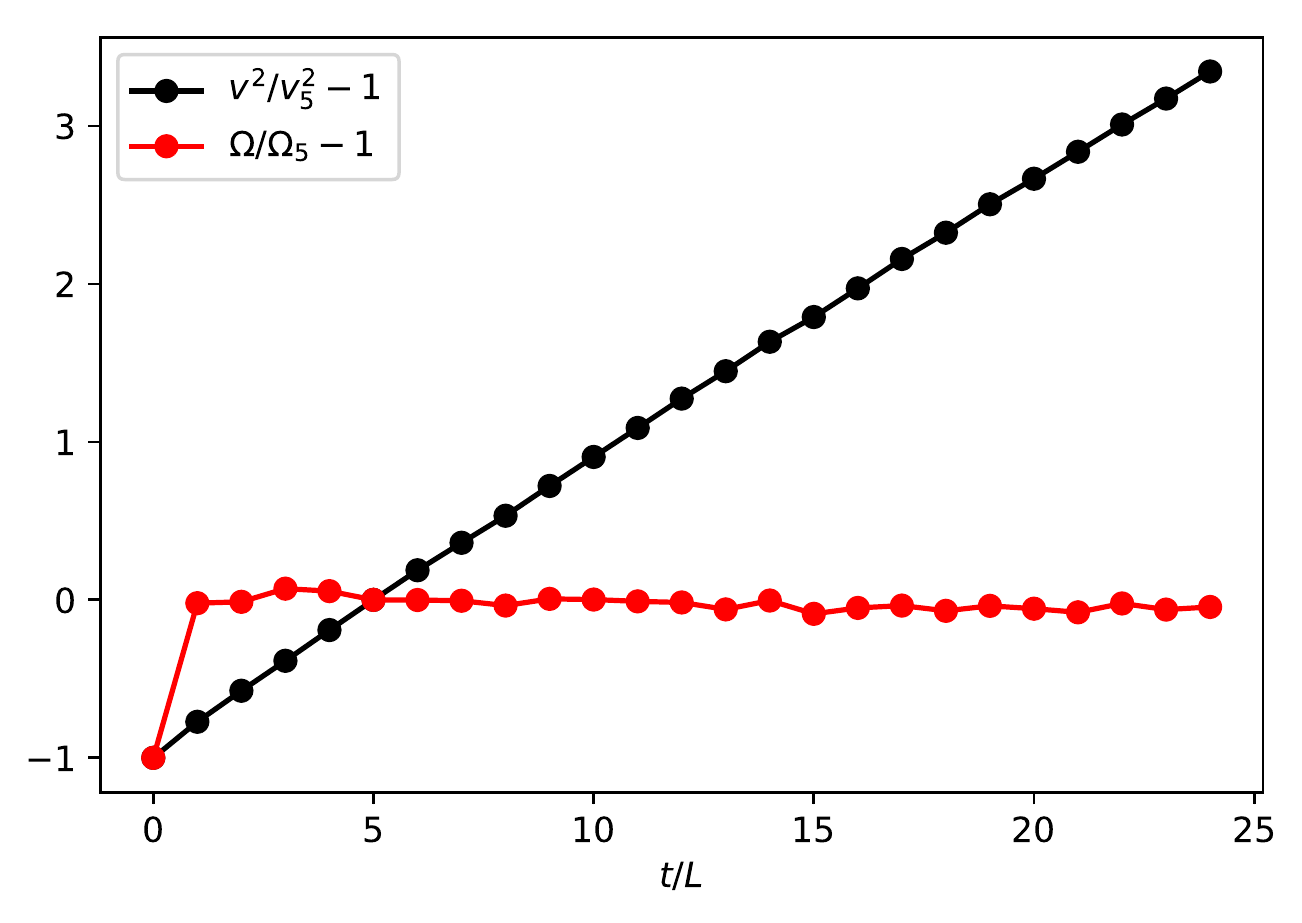}\\
\caption{Characteristics of the $F_i$-driven vorticity injection for a single realisation of a turbulent black brane horizon. Black points show the growth of kinetic energy with time, normalised by the kinetic energy at $t=5L$. This is consistent with a key assumption of the K41 dimensional analysis: that the relevant scale is a constant kinetic energy per unit time, $\varepsilon$. As further support we also show the approximate constancy of the total enstrophy of the system in red. \label{fig:characteristics}}
\end{center}
\end{figure}

So far we have provided evidence for the presence of K41 scaling in the form of power spectra and structure functions. Finally, we provide further support for why K41 predictions are seen here. A crucial part of the K41 analysis is that there is a dominant scale in the problem, $\varepsilon$, the rate of energy transfer, which is taken to be a constant. This governs how quickly kinetic energy is passed between vortices of different scales. Given that our inertial range is growing over time, exhibiting an indirect cascade from the injected $k$-scale downwards, we must be providing a constant injection of kinetic energy via the $F_i$ forcing. This is indeed the case, as is clearly shown in figure \ref{fig:characteristics}. We see a constant growth of $v^2$ with time (integrated over the torus), whilst the total enstrophy ($\Omega = \int d^2x \omega^2$) remains approximately constant.\footnote{For a discussion of the approximate constancy of $\Omega$ for \emph{unforced} equations see \cite{Rozali:2017bll}. Note that here we are forcing the equations and the approximate constancy of $\Omega$ emerges from the detailed dynamics.}

\section{Turbulent wakes}
\begin{figure*}[t]
\begin{center}
\includegraphics[width=\linewidth]{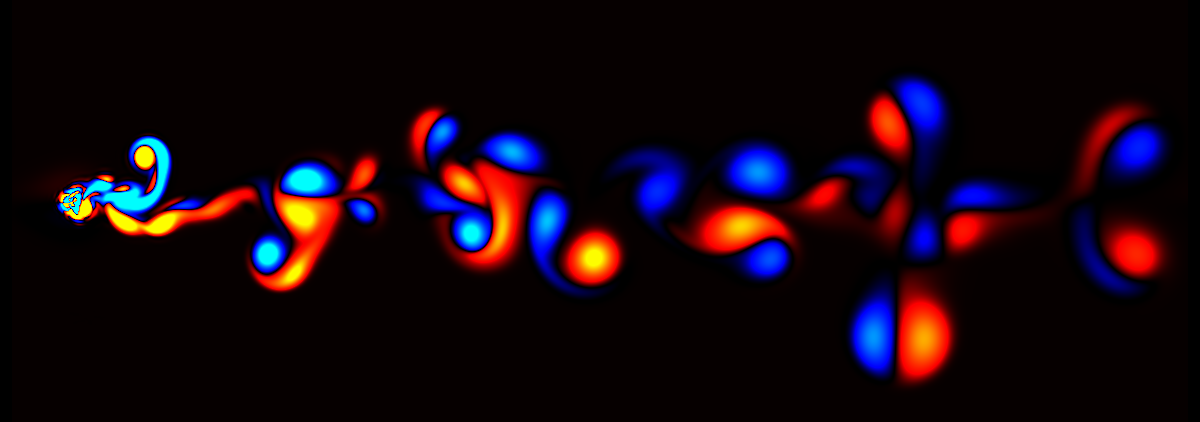}\\
\end{center}
\caption{Fixed-time snapshot of a turbulent wake of a gravitationally induced deformation of a large-$D$ black brane horizon in AdS. In this picture the fluid flows from left to right, past a circularly-symmetric Gaussian deformation at the left hand side, which was quenched from zero amplitude. This deformation corresponds to quenching a gravitational potential at the boundary of AdS. At late times the system will equilibrate as the fluid relaxes to the rest frame of the deformation. Colour indicates the vorticity. \label{fig:flyby}}
\end{figure*}
So far in this work we have examined universal turbulent dynamics of black hole horizons by explicitly adding a set of custom forcing terms $F_i$ to the Einstein equations which directly source vorticity.
Whilst such terms are the most convenient for achieving homogeneous isotropic turbulence and comparing to K41, it is desirable to have a fully gravitational realisation of turbulent dynamics. In fact our setup is capable to also address this question directly. 
To this end, in this section we consider the case where the forcing is instead provided by a gravitational deformation of the AdS conformal boundary metric, $\gamma_{tt}(t,\vec{x})$. The associated equations for such deformations simply result in a forcing term $F_i = a \nabla_i \gamma_{tt}/2$ as in \eqref{evo_p}.  We emphasize, however, that this means that the forcing term is entirely physical, and results from an explicit gravitational source. The fact that we can implement such a scheme is another illustration of the power of our setup for the study of gravitational turbulence. 

For this simulation we set initial conditions where the fluid is in uniform motion, and then quench $\gamma_{tt}$ from zero to a symmetric Gaussian profile at a fixed location $\vec{x}$ on the torus. The resultant dynamics, shown in figure \ref{fig:flyby}, clearly demonstrates that this deformation develops a turbulent gravitational wake on the horizon. Note that this example is not statistically homogeneous nor isotropic, and moreover it is decaying over time, and so we do not expect it to fall into the class of turbulent flows meeting the basic requirements to be described by K41.\footnote{It may be possible to utilise the other source terms $\gamma_{ij}, \zeta_i$ in \cite{Andrade:2018zeb} to directly source vorticity, though not for $\gamma_{ij}$ in $p=2$.}

We note that since drag force varies qualitatively between turbulent and laminar wakes we expect that this tidally-induced turbulence can impact the dynamics of strong-field gravitational processes. For example, the mechanism we have proposed may be of relevance to the dynamics of quark-gluon plasmas via AdS/CFT. To add further relevance to this scenario, the $\gamma_{tt}$ calculation we have performed may be taken in the same spirit as studies of near-extremal Kerr black holes and their near-horizon regions using CFT \cite{Guica:2008mu}. There, when a massive body falls into the near horizon region it appears as a source term in the CFT, as pointed out in \cite{Porfyriadis:2014fja}.  In our case, the $\gamma_{tt}$ may be viewed as the gravitational deformation due to such a massive body, and the appearance of a turbulent wake in this context may affect the plunge dynamics and associated waveforms for BH-BH or BH-NS mergers. Of course the extrapolation of our results to asymptotically flat spacetimes is not straightforward, particularly with regards to identifying an appropriate hierarchy of scales, and a direct computation would be required to confirm its astrophysical relevance.

\section{Discussion}

In this paper we have worked at strictly infinite spacetime dimension, $D$, though we considered dynamics constrained to $2+1$ of them. There are however reasons to expect that results we obtained continue to hold in lower dimensions. First, the dimensional analysis behind K41 scaling is dimension independent; indeed, $k^{-5/3}$ is predicted and seen in a range of $2+1$ and $3+1$ dimensional scenarios, as well as $2+1$ dynamics of an infinite-dimensional system as studied here. This universality occurs despite the clear differences in the dynamics that underpin the cascades; in $3+1$ the cascade is direct, whilst in $2+1$ there is also an inverse cascade. Thus whilst the detailed dynamics may change substantially as we lower the number of dimensions, we anticipate that K41 remains robust. Second, the map between gravity and hydrodynamics holds in all spacetime dimensions D.

We also highlighted the large-$D$ equations as a potentially useful model in the study of turbulence in general. These equations are simultaneously dissipative and exact in gradients, by virtue of the \emph{parametric} control afforded by the $1/D$ expansion. This should be contrasted with the usual treatment of hydrodynamics in a dynamical setting, where one typically truncates at a finite order and treats the resulting system of equations as exact -- a procedure which fundamentally changes the theory. As a consequence of this change one discovers physically undesirable qualities such as instabilities and acausal behaviour both for relativistic theories \cite{PhysRevD.31.725}, and non-relativistic theories \cite{Poovuttikul:2019ckt}. Furthermore, one must of course also verify post-hoc that the solution remained a good approximation within the framework of a perturbative gradient expansion. None of these concerns apply to our setup.

Finally we emphasise that whilst the large-$D$ equations of motion appear fluid-like, these are the radial \emph{constraint} pieces of a full solution to the Einstein equations. The radial \emph{evolution} equations were solved analytically as part of the construction, and so our solutions correspond to full black hole solutions to the Einstein equations. Thus the K41 behaviour we illustrate here from the fluid perspective is naturally encoded -- through the $a,p^i$ variables -- as geometric data in these spacetimes. In a more general setting the appropriate geometric data corresponding to the fluid observables may be difficult to identify, in which case it may be helpful to consider approaches to visualising horizon vorticity, for example \cite{Tendexes}.

\section*{Acknowledgments}
We thank Nils Andersson, Roberto Emparan, Ian Hawke, Achilleas Porfyriadis and Amos Yarom for useful discussion and comments on a draft of this paper.
This work has been supported by the Fonds National Suisse de la Recherche Scientifique (Schweizerischer Nationalfonds zur F\"orderung der wissenschaftlichen Forschung) through Project Grants 200021$\_$162796 and 200020$\_$182513 as well as the NCCR 51NF40-141869 The Mathematics of Physics (SwissMAP).
The work of T.A. is supported by the ERC Advanced Grant GravBHs-692951.
C.P. is supported by the European Unions Horizon 2020 research and innovation programme under the Marie Sklodowska-Curie grant agreement HoloLif No 838644. 
BW is supported by a Royal Society University Research Fellowship. 
Computations were performed at University of Geneva on the Baobab cluster.

\appendix
\section{Numerical Method\label{sec:numerics}}
The equations we are solving in this case are given by \eqref{evo_a} and \eqref{evo_p} with the right hand side of \eqref{evo_p} supplemented by the additional term
\be
F^i = \epsilon^{ij} \partial_j F.
\ee
Notice that this enters the equations in such a way to produce vorticity without directly sourcing $a$.
We consider the system on a torus with domain $[0,L]\times[0,L]$.
The force $F$ is constructed in such a way as to drive the system isotropically at a fixed energy scale, $m$, and consists of random combinations of all the Fourier amplitude coefficients whose corresponding wave vectors lie close to a circle of fixed radius in wave vector space. Specifically, given a discretisation scheme for the spatial directions, $F$ is given by
\be
F(t,x,y)= {\cal A} \sum_{i}^M {c^{(i)}(t) \cos \left( \frac{2\pi}{L}\vec k^{(i)} \cdot \vec x + \phi^{(i)}(t) \right) }
\ee
where $\vec k^{i}$ is a set of $M$ vectors in Fourier space that was sampled over an annular region $|\vec k^{(i)}|=m\pm \delta m$ in an isotropic way. The coefficient $\cal A$ is a fixed overall amplitude controlling the strength of the forcing. At times that are integer multiples of $\Delta t$ the mode coefficients $c^{(i)}$ are drawn from a normal distribution with zero mean and variance $\Delta t$, normalised such that
 \be
 \sum_i^M c^{(i)2}=1,
 \ee
while the angles $\phi^{(i)} $ are drawn from a uniform distribution. These random variables are assigned at times that are integer multiples of an interval $\Delta t$. For the times in between, i.e. from $t_1 = n \Delta t$ for some $n\in \mathbb{Z}^+$ and $t_2 = t_1 + \Delta t$ the forcing function is interpolated,
\be
F(t) = \cos\left(\frac{\pi (t-t_1)}{2 \Delta t}\right) F(t_1) + \sin\left(\frac{\pi (t-t_1)}{2 \Delta t}\right) F(t_2)
\ee
for $t_1\leq t<t_2$. This choice ensures the square-normalisation is maintained during the interpolation, with uncorrelated cross-terms cancelling after averaging.
We typically used $\Delta t= 10 \delta t$, where $\delta t$ is the time step used for the numerical evolution. 

A couple of comments are now in order. First of all, the benefit of forcing the system at a particular energy scale, $m$, is that there will be a sharp peak on the power spectrum at that scale, while the remaining of the spectrum will not be contaminated, allowing easier identification of scaling behaviours at larger and smaller energy scales. Furthermore, interpolating over the random amplitudes in this way allows us to use deterministic time-stepping techniques instead of stochastic ones.

Let us now move on to discuss the details of the numerical method used for solving these equations. We utilise a uniformly spaced discretisation of $x,y$ with $N_x, N_y$ grid points respectively, taking $N_x = N_y=N$.  We adopt fourth-order finite difference approximations of the derivative operators. The variables $a, p_x$ and $p_y$ are then evolved forward in time using fourth-order Runge-Kutta time-stepping (RK4), subject to periodic boundary conditions. 
In addition we add a sixth order Kreiss-Oliger dissipation term \cite{KO}. This is done by replacing the time derivative operator with 
\be
\partial_t\to\partial_t+\eta \frac{h^5}{64}(\partial_x^6+\partial_y^6)\,,\label{KOterm}
\ee
where $h=L/N$ is the grid spacing. Note that this term approaches to zero faster than the error in the spatial finite differences as $h\to 0$. At the highest resolutions considered, $N=1024$, we have also performed numerical evolutions with $\eta = 0$. The difference between $\eta =0$ and $\eta\neq 0$ is only visible near the UV in the power spectrum, as expected. The properties of the solutions away from the UV are not affected by $\eta$.

The last piece of information needed  for the time evolution is to specify the initial conditions. For simplicity, we consider homogeneous initial conditions for the evolution functions, namely $a=2\,, p_i=0$.

In the particular evolutions discussed we considered $N=1024, M=100, \eta=0.4,\delta t=6.4\times 10^{-4} L, {\cal A}= 0.005$, and $m=100$. In order to observe turbulence, the associated Reynolds numbers, $Re\sim L$, should be large, and in our simulations we use $L = 2\times10^5$. Random numbers are generated using the Mersenne twister algorithm \cite{MT}, and normally distributed variables are obtained using a Box-Muller transform \cite{box1958}. We work with double floating-point precision.

\section{Overview of GPU implementation}
Given the relative simplicity of the explicit update steps we have found the use of GPUs to be of significant utility. The methods used are standard and do not warrant a detailed exposition, however a high-level overview of the step-by-step procedure used may be of value. This is given below.

Seed the pseudo-random number generator uniquely for each run.
Allocate memory on the device to store, for all variables $a, p_x, p_y$: their spatial derivatives, first time derivatives, and their values at four RK4 intermediate steps. Allocate memory to store 100 pairs of $k_x, k_y$ values specifying the location of driving points in momentum space, as well as two buffers for the associated amplitudes and phases (two buffers are required because we interpolate between two sets of random variables over time). Initialise the allocated memory with initial data for the run, the driving point locations, initial random amplitudes and phases (all generated on the CPU).  Then, for each complete time step:
\begin{enumerate}[leftmargin=1em]
	\item Check if the random amplitudes and phases need updating at this time step. If so, cycle between the two buffers, compute new random values on the CPU and copy them to the device.
    \item For each RK4 intermediate step:
        \begin{itemize}[leftmargin=1em]
        \item Compute spatial derivatives in the $x$-direction (of the intermediate RK4 variable appropriate for this step). This is performed by splitting the $N\times N$ grid into $N$ thread blocks, one for each $y$-value, and using $N$ threads in each thread block to perform the computations at each $x$. In this way we can set the shared memory for each thread block to be the entire $y$-column, extended to include an extra 8 ghost points for periodicity. See, for example, \cite{nvidia}.
        \item Transpose and repeat for all $y$-derivatives (mixed second derivatives do not appear in the equations).
        \item Compute time derivatives using the previously computed values through the equations of motion.
        \item Add $F_i$ terms to the time derivatives as required. These are constructed from interpolating between $F_i$ values evaluated using initial and final amplitude/phase buffers.
        \item Fill the next RK4 intermediate buffer using the time derivatives. For the final RK4 intermediate step, the next RK4 buffer written is the first RK4 buffer.
        \end{itemize}
    \item Rarely, copy the field values (i.e. the first RK4 buffer) from the device and write to disk. In practise we found it invaluable to use the visualisation software \verb!VisIt! \cite{HPV:VisIt} and so we also write an appropriate `brick of values' descriptor file to accompany each binary data file.
\end{enumerate}
In this way the bulk of the evolution is performed on the device itself. The exception to this is also the slowest part of this procedure, namely the device synchronisation bottlenecks for step 1. 
\bibliographystyle{utphys}
\bibliography{turbs}{}

\end{document}